\documentclass[aps,prl,twocolumn,showpacs]{revtex4-1}
\usepackage[english]{babel}
\usepackage{amsmath,graphicx,bbm,color}
\usepackage{times}

\newcommand{\mathe}{\mathrm{e}}
\newcommand{\nobracket}{}
\newcommand{\tmem}[1]{{\em #1\/}}
\newcommand{\tmop}[1]{\ensuremath{\operatorname{#1}}}
\newcommand{\tmtextbf}[1]{{\bfseries{#1}}}
\newcommand{\STErev}[1]{{#1}}
\newcommand{\Brev}[1]{{#1}}
\def\SE{{\rm SE}}
\def\E{{\rm E}}
\def\S{{\rm S}}
\def\P{\Pi}

\begin{document}

\title{Two-step procedure to discriminate discordant from classical
correlated or factorized states}

\author{Simone Cialdi $^{1,2}$, Andrea Smirne $^{3,4}$, Matteo G.A. Paris $^{1}$, Stefano Olivares $^1$, Bassano Vacchini $^{1,2}$}

\affiliation{$^1$ \mbox{Dipartimento di Fisica, Universit{\`a} degli Studi di Milano, Via Celoria 16, I-20133 Milan, Italy}\\
$^2$ \mbox{Istituto Nazionale di Fisica Nucleare, Sezione di Milano, Via Celoria 16, I-20133 Milan,
  Italy}\\
$^3$ \mbox{Department of Physics, University of Trieste, Strada Costiera 11, I-34151 Trieste, Italy}\\
$^4$ \mbox{Istituto Nazionale di Fisica Nucleare, Sezione di Trieste,
  Via Valerio 2, I-34127 Trieste, Italy}}

  \begin{abstract}
    We devise and experimentally realize a procedure capable to detect
    and distinguish quantum discord and classical correlations, as
    well the presence of factorized states in a joint system-environment
    setting. Our scheme builds on recent theoretical results showing
    how the distinguishability between two reduced states of a quantum
    system in a bipartite setting can convey
    important information about the correlations present in the
    bipartite state and the interaction between the subsystems. The
    two addressed subsystems are the polarization and spatial degrees
    of freedom of the signal beam generated by parametric
    downconversion, which are suitably prepared by the idler
    beam. Different global and local operations allow for the
    detection of different correlations by studying via state
    tomography the trace distance behavior between suitable
    polarization subsystem states.
  \end{abstract}

\date{\today}
\maketitle


The study of a bipartite system is an ever present theme which has led to
important advancements in the understanding of quantum mechanics, especially
when the two parties cannot be put on an equal footing. The prototypical
situation is a measurement interaction, in which the
interest is all on the side of the system, calling for tools and ideas
allowing for an ever improving description of such interactions
{\cite{Holevo1982}}. The theory of open quantum systems has provided a natural
extension of these efforts, in which the quantum features of the measurement
apparatus are put in major evidence {\cite{Breuer2002}}, while
correlations between system and environment have received important attention
only more recently, thanks to the consolidation of quantum information theory
{\cite{Nielsen2000}}. The latest theoretical developments as well as
the refinement in the experimental techniques has led to a change of paradigm
in facing the system-environment ($\SE$) dynamics. The possibility has been envisaged
of actually exploiting the open quantum system, supposed to be liable to a
relatively easy and accurate experimental observation, as a quantum probe of
features of the environment, typically to be considered as a complex
system. Properties of the environment
which might be unveiled by an observation on the system up to now include the
detection of quantum phase transitions {\cite{Haikka2013b-Borrelli2013a-Gessner2014b}}, as
well as the assessment of correlations within the state of the environment
{\cite{Smirne2013a}}. These advancements have been based on the study in time of the distinguishability of different initial system
states \cite{Breuer2009b}, which has proven to be a fruitful strategy in
order to exploit a quantum
system as probe of features of a bipartite dynamics {\cite{Laine2010b,Smirne2011a,Cialdi2011a,Gessner2011a,Gessner2013a,Lanyon2013a,Gessner2014a}}.

In this paper we improve this approach
to devise a novel method for the determination of quantum
correlations, which play a crucial role both in quantum information
and in the development of quantum technologies. The approach is based on a two-step procedure, which
relying on measurements on the system only allows to determine whether a
given initial $\SE$ state actually contains quantum correlations,
as quantified by quantum discord or, if this is not the case, decide whether
it contains classical correlations or it is factorized. 
The relevance of this 
characterization lies in the fact that  
quantum discord has proven useful for different tasks in quantum
information processing (see e.g. \cite{discord2}).
Our
scheme goes beyond previous studies on the detection of initial correlations
{\cite{Laine2010b}} and of quantum discord {\cite{Gessner2011a}}, takes as
figure of merit for the distinguishability the trace distance among
statistical operators {\cite{Fuchs1999a}} and is experimentally realized in an
all-optical setup based on parametric downconversion (PDC) for the generation of
correlations {\cite{Cialdi2010a}}. At variance with other approaches,
relying on a measurement on multiple copies of the total
system \cite{Oh2011a-Aguilar2012a-Girolami2012a}, we here only perform tomographic measurements on one of the subsystems.

\begin{figure*}[htb]
  \includegraphics[width=0.9\textwidth]{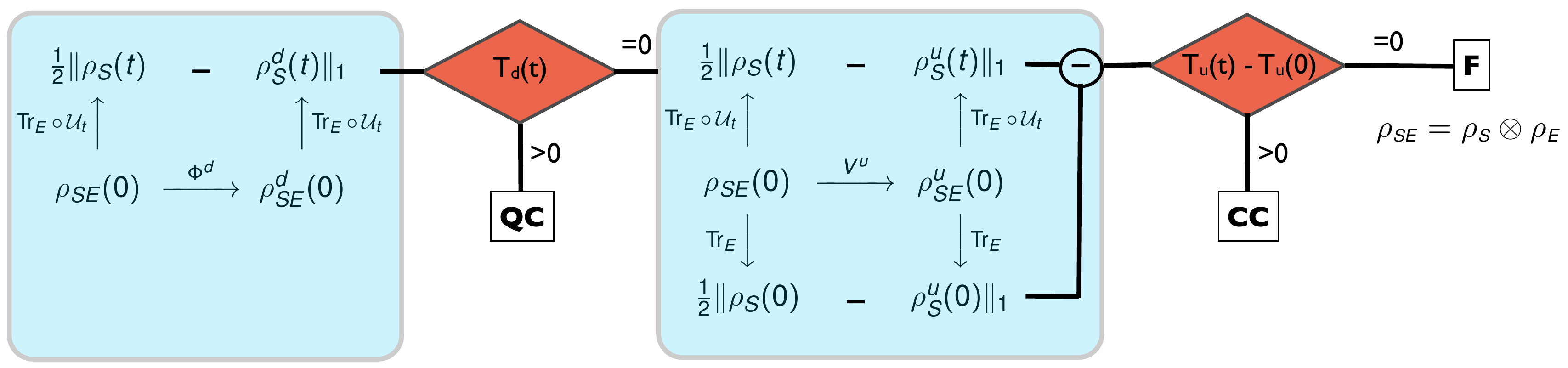}
\caption{\label{fig:logical}(Color online) Logical scheme of the cascading, \STErev{two-step} procedure
  exploited to discriminate among quantum correlated (QC), classically correlated (CC)
  or simply factorized (F) $\SE$ states. In the first stage (left box) a dephasing operation
  $\Phi^{d}$ which leaves invariant the marginals is applied, allowing to
  detect quantum correlations using as witness the trace distance $T_{d} ( t
  )$ between the reduced states evolved from original and dephased state. If
  no quantum correlations are detected, the second stage (right box) is entered, in which
  the growth in time of the trace distance of  initial states differing by a local unitary operation on the system $V^{u}$, i.e. $T_{u} ( t ) -T_{u} ( 0 )$,  allows to detect classical correlations or to conclude that the initial
  state is factorized. See the text for details.}
\end{figure*}
\paragraph{Detection of correlations}
We start considering a $\SE$ state, which might contain
correlations of some kind. For the experimental realization at hand we encode
the system in the polarization degrees of freedom of one of the beams in the PDC (referred to as the signal). The
environment corresponds to momentum (spatial) degrees of freedom of
the signal, while the other beam (usually referred to as the idler) is
exploited to prepare the initial state $\rho_{\SE} ( 0 )$.
In the first stage, the eigenstates of the reduced system state
$\rho_{\S}(0) = \mbox{Tr}_{\E}[\rho_{\SE} ( 0 )]$ are obtained by performing state tomography.
This allows us to define \Brev{the two orthogonal projections on the system eigenstates},
$\P$ and $\mathbbm{1} -\P$, \STErev{and to introduce a dephasing operation $\Phi^{d}$
such that  $\rho_{\SE} ( 0 ) \to \rho_{\tmop{\SE}}^{d} ( 0 ) \equiv
\Phi^{d} [ \rho_{\SE} ( 0 ) ]$, where
\begin{equation}
  \rho_{\SE}^{d} ( 0 ) = \P  \rho_{\SE} ( 0 ) \P
   + ( \mathbbm{1} -\P ) 
  \rho_{\SE} ( 0 ) ( \mathbbm{1} -\P ).
  \label{eq:deph}
\end{equation}
 The dephased state has the same marginals of the initial one but, according
to its expression, has zero quantum discord \cite{Ollivier2001a-Henderson2001a-Modi2012a}.}
As suggested in \cite{Luo2008a}, the difference
between $\rho_{\SE}^{d} ( 0 )$ and $\rho_{\SE} ( 0 )$,
as given by the trace distance,  provides a quantifier
of the quantum discord in the original state, namely:
\begin{equation}
  T = \frac{1}{2} \| \rho_{\SE}^{d} ( 0 ) - \rho_{\SE} ( 0 ) \|_{1} =
  \| \P  \rho_{\SE} ( 0 ) \P -\frac 12 \{\P, \rho_{\SE} ( 0 )\} \|_{1} .  \label{eq:lind}
\end{equation}
Now, one can prove the presence
of non-classical correlations in $\rho_{\SE} ( 0 )$ by just measuring
the system. In fact, if quantum correlations are present
the marginals of the system states after a time evolution $\mathcal{U}_{t}$
will generally differ, even if coinciding at the
initial time \cite{Gessner2011a}. This implies that the quantity
\begin{equation}
  T_{d} ( t ) = \frac{1}{2} \| \rho_{\S}^{d} ( t ) - \rho_{\S} ( t ) \|_{1}
  = \left\| \tmop{Tr}_{\E} \circ \,  \mathcal{U}_{t} [ \rho_{\SE}^{d} ( 0 ) - \rho_{\SE} ( 0 )  ] \right\|_1  \label{eq:wit}
\end{equation}
acts as a local witness of quantum discord in the initial state. This
witness is probabilistic in nature, since not every time evolution is
bound to reveal the existing quantum correlations.  However, as argued
in {\cite{Gessner2013a}}, the efficiency of the method is very high,
and in the case considered a fixed time evolution allows for the
detection of quantum discord in the whole range of states which can be
prepared, apart from a set of measure zero. In the general case it has
been shown that the average over the set of unitaries not only detects
the quantum discord, but also allows to quantify it. This first stage
of the detection scheme is described in the first section of the
logical scheme in Fig.~\ref{fig:logical}. If the witness
provided by the expression Eq.~(\ref{eq:wit}) is positive, then the
state $\rho_{\SE} ( 0 )$ does contain quantum correlations
corresponding to non zero discord.  On the other hand, if $T_{d} ( t )
= 0$,
then the second stage of the cascading
procedure is entered \STErev{(second section in
Fig.~\ref{fig:logical})}. At this level we have already checked the absence of
quantum correlations,  \STErev{therefore we should perform a measurement
involving only the system to check whether $\rho_{\SE}(0)$ is a factorized state or
contains classical correlations. Also in this case} the presence of initial
correlations can be unveiled by a growth of the trace distance between
different initial states above the initial value as a consequence of the
$\SE$ time evolution {\cite{Laine2010b}}: while
the considered condition is in principle only sufficient, the considered time
evolution allows to detect with unit efficiency the considered class of
states. In order to generate another initial $\SE$ state without
introducing correlations we perform a \STErev{local} unitary transformation denoted by
$V^{u}$, which only affects the system degrees of freedom. Given the fact that
the marginal states of the environment are left \STErev{unchanged} by \STErev{$V^{u}$},
the growth of the trace distance indeed witnesses the presence of
initial correlations, rather than of different initial environmental states.
We are then led to consider the behavior of the trace distance between the
reduced system state $\rho_{\S}$ and \STErev{its transformed counterpart} $\rho^{u}_{\S}$
at the initial and at a later time. If the difference
\begin{equation}
  T_{u} ( t ) -T_{u} ( 0 ) = \frac{1}{2} \| \rho_{\S}^{u} ( t ) - \rho_{\S}
  ( t ) \|_1 - \frac{1}{2} \| \rho_{\S}^{u} ( 0 ) - \rho_{\S} ( 0 ) \|_1
  \label{eq:ic}
\end{equation}
which acts as correlation witness is greater than zero, then $\rho_{\SE} ( 0 )$ has
classical correlations, otherwise the state is actually factorized, i.e.
$\rho_{\SE} ( 0 ) = \rho_{\S} ( 0 ) \otimes \rho_{\E} ( 0 )$ (see Fig.~\ref{fig:logical}).

\begin{figure}
  \includegraphics[width=.45\textwidth]{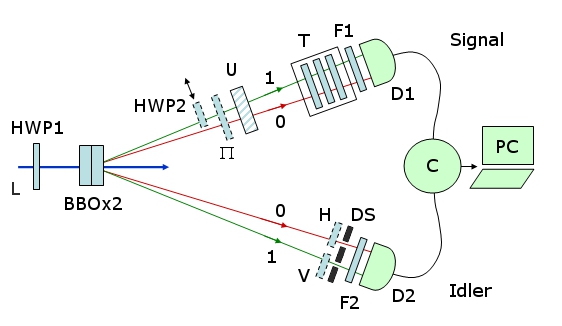}
  \vspace{-0.3cm}
  \caption{\label{fig:schema}(Color online) Scheme of the apparatus.  L is the pump laser, 
  HWP1 and HWP2 are half-wave plates, BBOx2 two BBO crystals,
  $\P$ a polarizer and  U is the  spatial light modulator. T is the tomographic scheme
  (a quarter-wave plate, a HWP and a polarizer), H and V denote two polarizers
  aligned along the horizontal and vertical axes, respectively, DS is the
  double slit, F1 a high pass filter (780~nm), F2 an interference filter
  (bandwidth of 10~nm and central wavelength of
810~nm); D1, D2 are detectors and C the coincidences counter.
The components drawn with a dashed line have to be modified or moved
according to the different stages of the procedure.}
\end{figure}
\paragraph{Experimental realization}
In our experiment $\SE$ states with different correlations have
been generated, and the two-step procedure
described above for the discrimination of correlations has been tested,
providing in particular an experimental verification of the scheme for the detection
of quantum discord proposed in {\cite{Gessner2011a}}.
\par
\begin{figure}
\hspace{-1em}
\includegraphics[width=.24\textwidth]{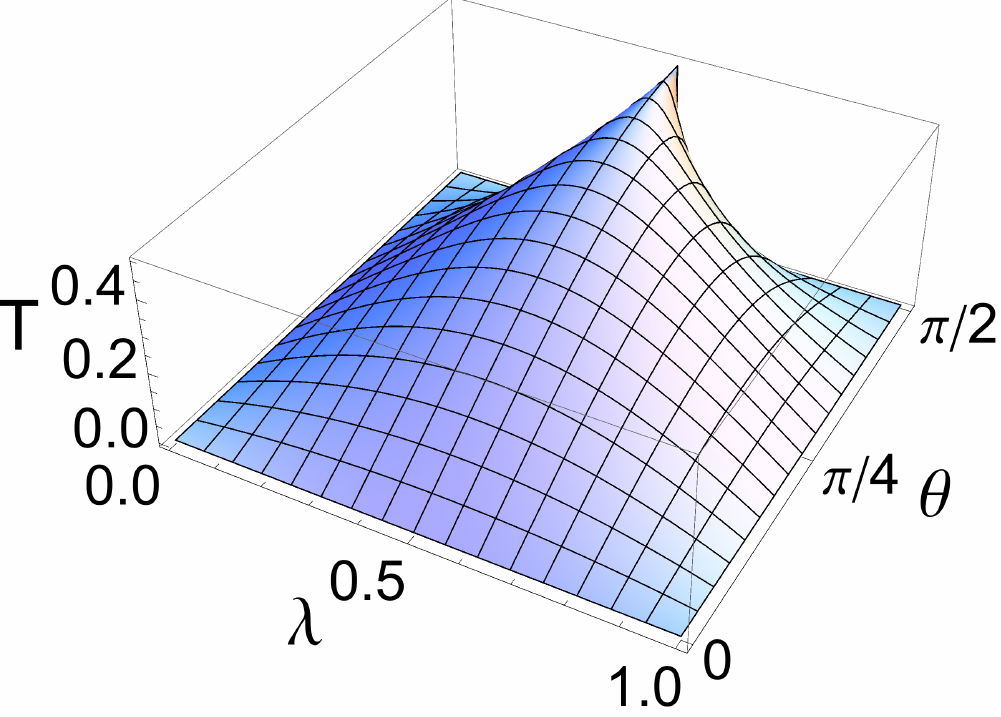}
\hfill
\includegraphics[width=.24\textwidth]{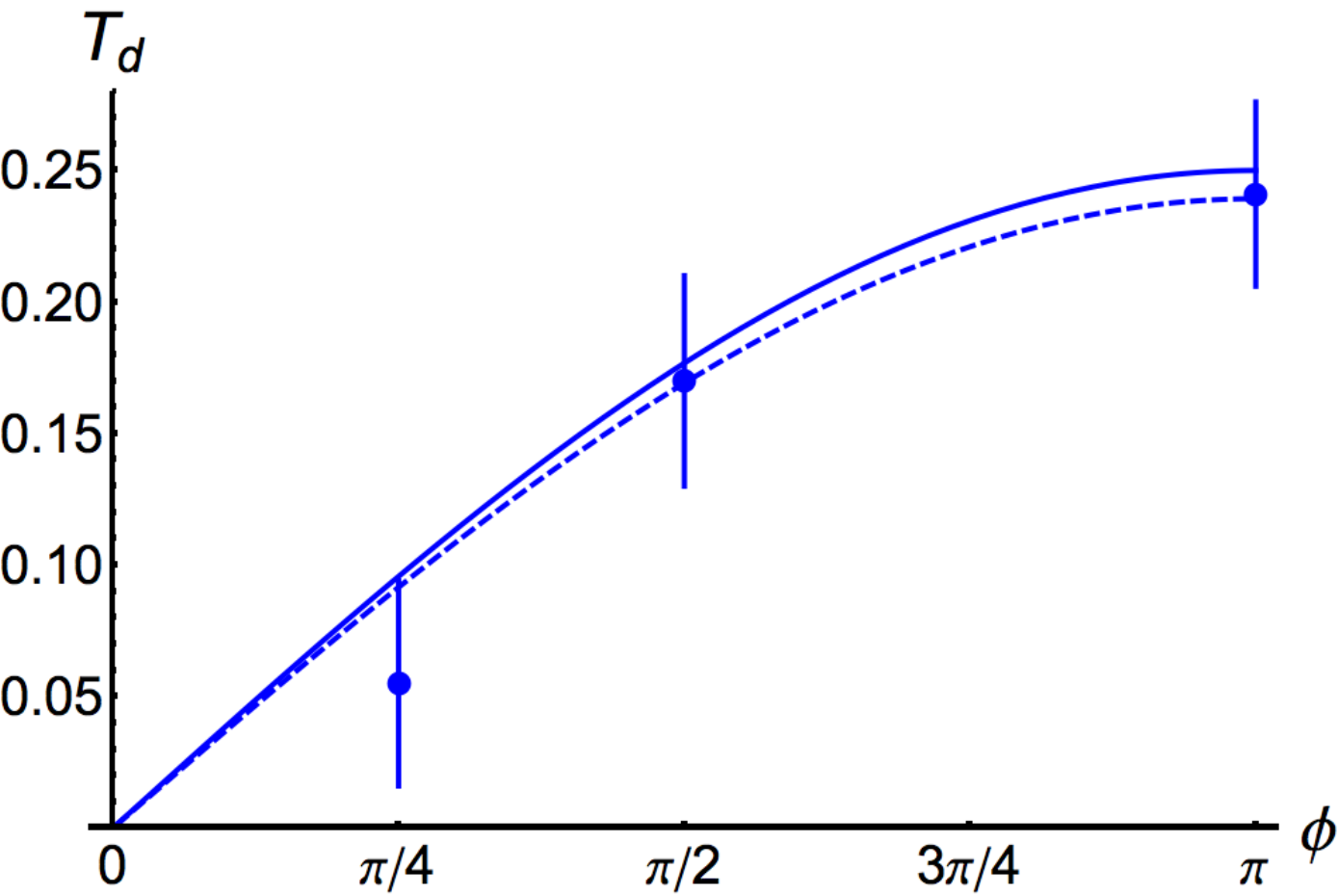}
  \caption{\label{fig:premise}(Color online) (Left) The amount of quantum discord 
  $T$ defined in Eq.~(\ref{eq:lind}) for a state given in Eq.~(\ref{eq:qc}), as a function
  of  $\lambda$ and $\theta$. (Right) \STErev{Measured values of $T_d$
  for different values of the parameter $\phi$,
  which describes the interaction between system and environment}. 
  The two points for $\phi=\pi/2$ and $\phi=\pi/4$ have been obtained for
  $\lambda=0.5$, whereas for $\phi=\pi$ we have set $\lambda=0.48$.
  The solid
  and dashed lines correspond to $\lambda =0.5$ and $\lambda =0.48$, respectively.}
\end{figure}

Our experimental apparatus, sketched in Fig.~\ref{fig:schema}, is based on PDC
generated by two 1~mm adjacent type-I Beta-Barium Borate
(BBO) crystals, oriented with their optical axes aligned in perpendicular
planes and pumped by a 10~mW, 405~nm cw diode laser (Newport LQC405-40P). The two
BBO crystals generate \STErev{the signal and idler} states with perpendicular polarization and the interference
filter (F2) ensures a good spatial correlation between signal and idler
{\cite{Cialdi2011b,Smirne2013a,Cialdi2012a}}. We generate two channels 0 and 1
\STErev{(corresponding to the momentum states $| 0 \rangle$ and $| 1 \rangle$, respectively)}
with a double slit (DS) positioned along the idler path. This scheme
allows us \STErev{to act on the idler beam to prepare the signal beam in} the three cases
of interest and to easily control and change the amplitude of the
polarizations. The arrangement of the two BBO crystals
produces a factorized state between polarization and momentum, namely $\rho_{p}
\otimes \rho_{m}$. Both components are generally described by a mixture of the
form {\cite{Smirne2011a}} $\rho_{k} = P_{k} \rho^{\tmop{ent}}_{k} + ( 1-P_{k} )
  \rho^{\tmop{mix}}_{k}$,
where $k=p,m$, the statistical operator $\rho^{\tmop{ent}}_{k} = | \psi_{k}
\rangle \langle \psi_{k} |$ denotes a pure entangled state and
$\rho^{\tmop{mix}}_{k}$ the corresponding mixed \STErev{counterpart}. The weight $P_{k}$ is
naturally interpreted as purity of the state, but does not play a role in the
present treatment which studies the correlations between polarization and
momentum. The states for polarization and momentum read
$| \psi_{p}\rangle = \sqrt{\lambda} |HH
\rangle + \sqrt{1- \lambda} |VV \rangle$ and
$| \psi_{m} \rangle = \frac{1}{\sqrt{2}} ( |00 \rangle +|11 \rangle )$
respectively, where $H$ and $V$ denote
horizontal and vertical polarizations. The relative weight of the two
polarization states parametrized by $0\le \lambda \le 1$ can be adjusted at will by means
of a half-wave plate (HWP) located in the path of the pump laser, while the
balance in the momentum degrees of freedom is obtained by a careful alignment
of the preparation apparatus and optimizing the phase-matching between the
crystals.
Controlled correlations between system \STErev{(polarization)} and environment
\STErev{(momentum)} can be introduced
inserting in the idler beam a horizontal and a vertical polarizer in the paths
corresponding to the momenta denoted by $0$ and $1$ respectively. If no
further operation is performed, the obtained state is of the form
\begin{equation}
  \rho^{\tmop{CC}}_{\SE}  =  \lambda | H \rangle \langle H | \otimes | 0
  \rangle \langle 0 | + ( 1- \lambda ) | V \rangle \langle V | \otimes | 1
  \rangle \langle 1 | ,  \label{eq:cc}
\end{equation}
which clearly exhibits only classical correlations, while states with non zero quantum
discord are generated by inserting a half-wave plate (HPW2) in the momentum
channel $1$ of the signal beam, thus obtaining
\begin{equation}
  \rho^{\tmop{QC}}_{\SE} = \lambda | H \rangle \langle H | \otimes | 0
  \rangle \langle 0 | + ( 1- \lambda ) | \theta \rangle \langle \theta |
  \otimes | 1 \rangle \langle 1 | ,  \label{eq:qc}
\end{equation}
where $| \theta \rangle = \cos ( \theta ) | H
\rangle + \sin ( \theta ) | V \rangle \nobracket$. \STErev{In the left panel of
Fig.~\ref{fig:premise}} we plot the quantum discord in such a state as
quantified by Eq.~(\ref{eq:lind}). The absence of polarizers in the idler path
leads to take the trace over the idler degrees of freedom and therefore to the
factorized state
\begin{equation}
  \rho^{{\rm F}}_{\SE} = ( \lambda | H \rangle \langle H | + ( 1- \lambda ) | V
  \rangle \langle V | ) \otimes \frac{1}{2} ( | 0 \rangle \langle 0 | + | 1
  \rangle \langle 1 | ).  \label{eq:f}
\end{equation}
The eigenstates of the reduced system states, whose knowledge is necessary to determine the dephasing operation described in Eq.~(\ref{eq:deph}),
that is the projections $\P$ and $\mathbbm{1}-\P$, 
are obtained through the full tomography (T) of the polarization states
{\cite{James2001a-Banaszek1999a}}, as depicted in
Fig.~\ref{fig:schema}. \STErev{The projections are implemented by means of polarizers according
to the measured eigenstates.} The interaction between system and environment
is obtained by a spatial light modulator (U), which can insert a position and
polarization sensitive phase in the signal. In particular we have realized an
evolution corresponding to a phase-gate, acting on the momentum corresponding
to channel $1$ only by applying a phase to the polarization degrees of freedom
according to the operator $\tmop{Diag} ( \mathe^{i \phi} ,1 )$ in the $\{ |
\nobracket H \rangle , | \nobracket V \rangle \}$ basis. As shown in
\STErev{the right panel of 
Fig.~\ref{fig:premise}}, the optimal performance in the correlation detection
is obtained for $\phi = \pi$, which has thus been taken as reference
value.
In the following, the time specification ``$0$'' will denote the state right after the preparation,
while the time ``$t$''' will identify the state after the interaction.
The unitary transformation $V^{u}$ on the system degrees of freedom only, used
to prepare the other reference state for the second stage of the two-step
procedure of Fig.~\ref{fig:logical}, is obtained by inserting a half-wave plate intercepting both
momenta in the idler beam.
\begin{figure}
  \includegraphics[width=.45\textwidth]{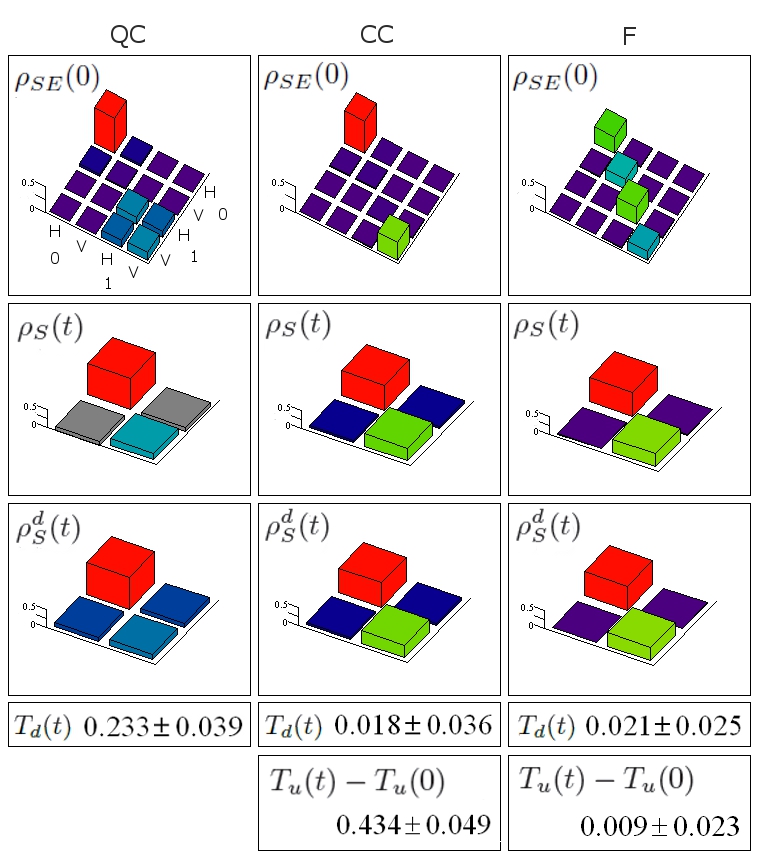}
  \caption{\label{fig:tomography} (Color online) Tomographic measurements of the states
  involved in the experiment. In the left column the case of a state of the
  form Eq.~(\ref{eq:qc}) with $\lambda =0.7$ and $\theta = \pi /4$ has been
  considered. From top to bottom we have plotted the observed values for
  $\rho_{\SE} ( 0 )$, $\rho_{\S} ( t )$ and $\rho_{\S}^{d} ( t )$ respectively.
  In the central column we provide the corresponding measurements for the
  state Eq.~(\ref{eq:cc}) with $\lambda =0.64$. The value $T_{d} ( t )$ of the
  trace distance Eq.~(\ref{eq:wit}) is here compatible with zero
  \Brev{according to the experimental error}, testifying
  the absence of quantum discord, while the positivity of $T_{u} ( t ) -T_{u}
  ( 0 )$ given by Eq.~(\ref{eq:ic}) shows the detection of classical
  correlations. In the right column the considered state corresponds to
  Eq.~(\ref{eq:f}) with $\lambda =0.65$, and the factorized structure of the
  state is unveiled by the value of $T_{u} ( t ) -T_{u} ( 0 )$, which is zero
  within the experimental value. The time specification 0 and $t$ denote the
  states right after the preparation and the interaction stages respectively.}
\end{figure}
\par
The experimental results are summarized in Fig.~\ref{fig:tomography},
that reports the data of the tomographic analysis.  In the first row
examples of the system-environment states corresponding to
Eqs.~(\ref{eq:qc}), (\ref{eq:cc}) and (\ref{eq:f}), respectively, are
considered for specific values of $\lambda$ and $\theta$.  From the
tomographic data we retrieve the expression for the dephasing
operation $\Phi^{d}$ to be implemented. In the second row the reduced
system states after the time evolution corresponding to a phase gate
are given, to be compared via trace distance with the reduced states
plotted in the third row and obtained by applying $\Phi^{d}$ to the
overall state before the evolution. The experimentally measured value
of the trace distance growth corresponding to Eq.~(\ref{eq:wit}) is
given in the fourth row. When this value is zero \STErev{(within the
  experimental errors)}, thus pointing to the absence of quantum
discord, a further analysis corresponding to the second stage of the
scheme in Fig.~\ref{fig:logical} is performed. Therefore, we first
apply a local unitary operation $V^{u}$ to the system and then measure
the quantity Eq.~(\ref{eq:ic}), whose positivity reveals the presence
of correlations in the initial state, as detected by a growth of the
distinguishability in time between different initial reduced system
states. The experimental values for the quantity in Eq.~(\ref{eq:ic})
are given in the last row of Fig.~\ref{fig:tomography}, showing that
indeed a factorized state can be detected within the experimental
accuracy. 
In fact the indistinguishability of two statistical operators
corresponding to zero trace distance can be consistently assessed within a
tomographic approach since quantum tomography is a statistically
reliable procedure, meaning that for any finite number of repeated
preparations one obtains an estimate with a predictable standard
deviation, thus leading to error bars following the standard
statistical scaling for any quantity evaluated using the reconstructed
density matrix \cite{noteTomo}.
\par
\begin{figure}
  \includegraphics[width=.4\textwidth]{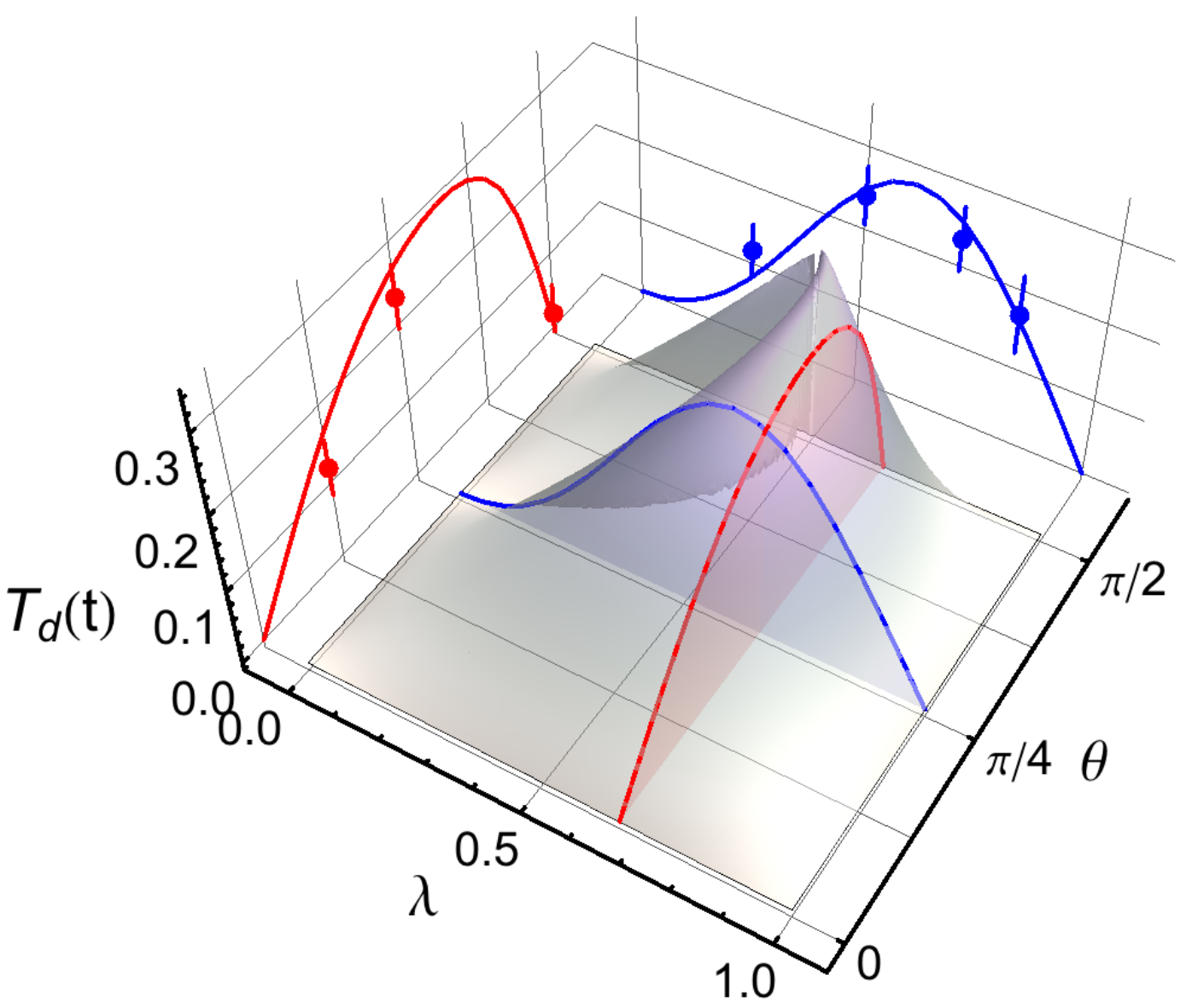}
  \caption{\label{fig:parameters} (Color online) Experimental results for the trace distance
  Eq.~(\ref{eq:wit}) corresponding to different values of the parameters
  $\lambda$ and $\theta$, as compared to the theoretical prediction given by
  the smooth surface. The red curve corresponds to $\lambda =0.65$, while the
  blue curve is fixed by $\theta = \pi /4$. The experimental points are
  plotted on the projected curves to improve their visibility.}
\end{figure}

The reliability of the method has been further tested by measuring the growth
of the trace distance between the dephased states after the interaction as
quantified by Eq.~(\ref{eq:wit}) for different values of $\lambda$ and
$\theta$ and comparing it with the theoretical prediction. The result is
plotted in Fig.~\ref{fig:parameters}, where different experimental points are
measured along lines with fixed relative weight $\lambda$ and varying angle
$\theta$, as well as vice-versa. The theoretical expression is given by the
smooth surface. As it appears the trace distance Eq.~(\ref{eq:wit}) lies above
zero, thus detecting the quantum discord of the state plotted in there left panel of
Fig.~\ref{fig:premise}, for all possible values of the parameters $\lambda$
and $\theta$, apart from a set of measure zero corresponding to the points on
the line $\lambda [ \cos ( 2 \theta ) -1 ] = \cos ( 2 \theta )$.

\paragraph{Conclusions and outlook}
We have suggested and demonstrated a simple all-optical setup to
detect and discriminate different kind of $\SE$ correlations by
performing measurements on the system only. The scheme consists of a two-step
procedure. At each step information about the presence and the nature of correlations
is extracted by tomographically estimating the distinguishability between
system states after the action of suitable global or local quantum
operations. In particular, we first assess the presence
of quantum discord as quantified by the measurement induced disturbance 
\cite{Luo2008a,Gessner2011a}, and then, in the absence of quantum discord, we further
determine the factorizability of the state versus the presence of classical
correlations, exploiting the connection between initial correlations and growth
of trace distance {\cite{Laine2010b}}. 
The successful realization of our procedure is based on the
implementation of a dephasing map on the $\SE$ state and on the reliable 
detection of quantum discord.
Our procedure can be easily adapted to different experimental settings, 
the basic requirement being the realization of the dephasing map and
the capability to perform state tomography
on the sole system. Our results pave the way for reliable detection and 
discrimination of environments or $\SE$ features in systems of interest
for quantum technology.
\paragraph*{Acknowledgments}
This work has been supported by MIUR (FIRB RBFR10YQ3H ``LiCHIS'').
BV and AS gratefully acknowledge financial support by the EU projects COST Action MP
1006 and NANOQUESTFIT respectively.

\end{document}